\documentclass[prl,twocolumn,floatfix,showpacs,showkeys,preprintnumbers,amssymb,amsmath]{revtex4}
\usepackage[latin1]{inputenc}
\usepackage{graphicx}
\usepackage{dcolumn}
\usepackage{bm}
\usepackage[mathscr]{eucal}
\usepackage{epsfig}
\usepackage{rotating}

\begin{document}
\newcommand{\beq}{\begin{equation}}
\newcommand{\eeq}{\end{equation}}
\newcommand{\ket}{\rangle}
\newcommand{\bra}{\langle}
\newcommand{\A}{\mathbf{A}}
\preprint{ }
\title{Optimal multiqubit operations for Josephson charge qubits}

\author{Antti\ O.\ Niskanen}
\altaffiliation[Currently at ]{VTT Information Technology, Microsensing, POB 1207, 02044 VTT, Finland}
\email{antti.niskanen@vtt.fi}
\author{Juha J. Vartiainen}
\email{juhav@focus.hut.fi}
\author{Martti\ M.\ Salomaa}
\affiliation{Materials Physics Laboratory, POB 2200 (Technical Physics), FIN-02015 HUT,
Helsinki University of Technology, Finland}

\date{\today}

\begin{abstract}
We introduce a method for finding the required control parameters
for a quantum computer that yields the desired quantum
algorithm without invoking elementary gates. We concentrate on the
Josephson charge-qubit model, but the scenario is readily extended
to other physical realizations. Our strategy is to
numerically find any desired double- or triple-qubit gate. The
motivation is the need to significantly
accelerate quantum algorithms in order to fight decoherence.
\end{abstract}



\maketitle

Quantum computing algorithms are realized through unitary
operators that result from the temporal evolution of the quantum
system under consideration. Typically, these are achieved with a
sequence of universal gates \cite{elementary} which act analogously
to the elementary gates of digital computers.
Quantum computers hold the promise of exponential speedup with respect
to classical computers owing to the massive parallelism arising from
the superposition of quantum bits, qubits; for introductions to quantum computing
and quantum information processing,  see Refs.~\cite{gruska}.
Several different physical implementations of quantum computing
have been suggested; in particular quantum computing
with Cooper pairs has  been proposed \cite{averin}.

Superconducting circuits \cite{schon}
feature controlled fabrication and scalability \cite{You}; their drawback is that the leads
inevitably couple the qubit to the environment, thereby introducing decoherence \cite{zurek}.
In a superconductor, the number of the Cooper pairs and the phase of
the wavefunction constitute conjugate variables. The majority of investigations
has focused either on the charge regime where the number of Cooper pairs
is well defined \cite{nakamura}, or on the flux regime where the phase is well defined \cite{pc}.
Qubits utilizing current-driven large Josephson junctions have been tested experimentally \cite{martinis}.
The decisive experimental progress reported in Ref.~\cite{vion} has demonstrated
that it is possible to realize $10^4$ elementary quantum gates with
Josephson-junction qubits. Here we consider Josephson charge qubits.

In this Letter we propose a method to construct arbitrary two- or
three-qubit quantum gates by solving the numerical optimization
problem of control parameters for a Josephson charge qubit register.
We show that it is possible to numerically find the required
control-parameter sequences even for nontrivial three-qubit gates
without employing elementary gates. Recently, it has been
suggested \cite{ours} how to solve a similar problem in the
context of holonomic quantum computation \cite{zanardi},
where time does not appear as an explicit parameter.
In the present context, the time evolution appears
through the Schr\"odinger equation.

\begin{figure}
\begin{picture}(130,70)
\put(-55,0){\includegraphics[width=0.45\textwidth]{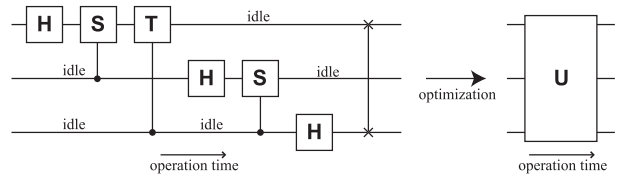}}
\end{picture}
\caption{\label{fg:principle} Instead of implementing the three-qubit quantum Fourier
transform with the help of elementary gates, we determine a
gate that performs the entire three-qubit operation with a single
control loop. Note that idle time is avoided.}
\end{figure}

The motivation underlying the investigation of this approach is
the need to overcome effects of decoherence. The implementation of
a quantum algorithm which is composed of elementary gates is rarely
optimal in execution time since the majority of qubits is most
of the time inactive, see Fig.~\ref{fg:principle}. The
decomposition into elementary gates works extremely well with
classical digital computers. However, in the context of quantum
computing the number of consecutive operations is strictly limited
by the short time window set by interactions with the environment.
It is therefore of prime importance to concentrate on the
implementations of quantum algorithms \cite{siewert,palao,wang}.
We consider the construction of quantum algorithms out of larger
building blocks. Whereas careful design and manufacturing can
significantly increase the decoherence time, our scenario can
serve to reduce the number of the operations needed.

The Josephson charge qubit utilizes the number degree of freedom
of a nanoscale Josephson-junction circuit. The states of the qubit
correspond to either zero or one extra Cooper pair residing on the
superconducting island, usually denoted by $|0\ket$ and $|1\ket$,
respectively. The Cooper pairs can tunnel coherently to a
superconducting electrode. The charging energy of the qubit can be
tuned with the help of an external gate voltage, whereas tunneling
between the states is controlled with the help of an external
magnetic flux.

The explicit single-qubit Hamiltonian for the qubit $i$ is
\beq\label{eq:single}
H^i_{\text{single}}=-\frac{1}{2}B^i_z\sigma_z-\frac{1}{2}B^i_x\sigma_x,
\eeq where the standard notation for Pauli matrices has been
utilized. Here $B_z^i$ is a tunable parameter which depends on the
gate voltage, while $B_x^i$ can be controlled with the help of a
flux through the SQUID. Note that setting $B_z^i=B_x^i=0$ results
in degeneracy. At the degeneracy point, there will be no change in
the physical state of the system. In the case of single-qubit
gates, it is easy to see from this model that any rotation in
$SU(2)$ can be performed on the qubits. Note that $U(2)$ is not
available since the Hamiltonian is traceless. In general, we
cannot achieve $U(2^N)$ for $N$ qubits since the Hamiltonian of
the entire quantum register turns out to be traceless. However,
the global phase factor is not physical since it corresponds to a
redefinition of the zero level of energy.

\begin{figure}
\begin{picture}(130,105)
\put(-40,0){\includegraphics[width=0.40\textwidth]{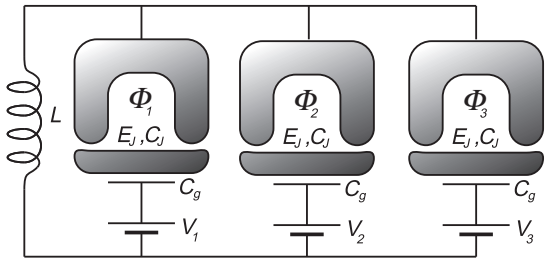}}
\end{picture}
\caption{\label{fg:qubits} Schematic illustration of three
Josephson charge qubits with inductive coupling. The adjustable
parameters include the gate voltages $V_i$ and the enclosed fluxes
$\Phi_i$.}
\end{figure}

Qubits can be coupled by connecting them in parallel to an
inductor, see Fig.~\ref{fg:qubits}. This scenario has the benefit
of allowing for a longer decoherence time and that of being
tunable. The resulting coupling term in the Hamiltonian between
the qubits $i$ and $j$ is then of the form \beq\label{eq:couple}
H_{\text{coupling}}=-CB^i_xB^j_x\sigma_y\otimes \sigma_y, \eeq
where $C$ is a positive parameter depending on the capacitances of
the qubits and also on the inductance. It follows from
Eqs.~(\ref{eq:single}--\ref{eq:couple}) that one can apply
nontrivial two-qubit operations by simultaneously turning on the
SQUIDs of the two qubits, although the $\sigma_x$-term will be
turned on as well. All the other qubits must have their SQUIDs
turned off. On the other hand, one-qubit $\sigma_x$-operations
require that all but one SQUID is turned on. Note that in the
present context it is actually impossible to perform independent
operations on any two subsets of the quantum register due to the
inductive coupling. Since one must also take into account the
decoherence mechanism, it is not practical to let most qubits
reside at their degeneracy point. The question arises whether it
would rather prove more efficient to try and find some scheme of finding
larger quantum operations, instead of using elementary gates.

To tackle the challenge posed above, we concentrate on finding
quantum gates numerically. The structure of the Josephson-qubit
Hamiltonian is such that it is not immediately transparent how one
would actually construct even the basic controlled-NOT gate. We
accomplish this by considering loops $\gamma(t)$ in the control-parameter
space spanned by $\{B_x^j(t)\}$ and $\{B_z^j(t)\}$.
Therefore, the function $\gamma(t)$ is of the vector form
\beq
\gamma(t)=\left[\begin{matrix}B_z^1(t) & \dots & B_z^N(t) &
B_x^1(t) & \dots & B_x^N(t)\end{matrix}\right]^T,
\eeq
where we have assumed a register of $N$ qubits. The temporal evolution
induces the unitary operator
\beq\label{eq:U}
U=\mathcal{T}
\exp\left(-i\int_{\gamma(t)} H(\gamma(t))dt\right),
\eeq
where $\mathcal{T}$ stands for the time-ordering operator and we choose
$\hbar=1$. The integration is performed along the path formed by
$\gamma(t)$ where the loop starts at the origin, i.e., at the
degeneracy point. We will restrict the path to a special class of
loops, which form polygons in the parameter space. Thus the
parameters vary in time at a piecewise constant speed, and none of
the parameters is turned on or off instantaneously. We further set
the time spent in traversing each edge of the polygon equal to
unity. This limitation could be relaxed, in which case the length
of each edge in time would be an additional free parameter. We
also set $C=1$ in Eq.~(\ref{eq:couple}). This can be achieved by
properly fabricating the inductor, but we have every reason to
believe that the algorithm will work for other choices of $C$ as
well. Hence, in order to evaluate Eq.~(\ref{eq:U}) one only needs
to specify the coordinates of the vertices of the polygon, which
we denote collectively as $X_\gamma$.  Numerically, it is easy to
evaluate the unitary operator in a stable manner by further dividing the
loop $\gamma(t)$ into tiny intervals that take the time
$\Delta t$ to traverse. If $\gamma_i$ denotes all the values of
the parameters in the midpoint of the $i^{th}$ interval, and $m$
is the number of such intervals, then we find to a good
approximation
\beq
U_{X_\gamma}\approx \exp(-iH(\gamma_m)\Delta t)\ldots \exp(-iH(\gamma_1)\Delta t).
\eeq
We now proceed to
transform the problem of finding the desired unitary operator into
an optimization task. Namely, any $\hat{U}$ can be found as the
solution of the problem of minimizing the error functional
\beq
\label{eq:error} f(X_\gamma)=\|\hat{U}-U_{X_\gamma}\|_F
\eeq over
all possible values of $X_\gamma$. Here $\|\cdot\|_F$ is the
Frobenius trace norm defined as
$\|{A}\|_F=\sqrt{\mathrm{Tr}\left({A}^\dagger{A}\right)}$. The
number of adjustable vertices of the polygon $\nu$ is kept fixed
from the beginning. One needs to have enough vertices to
parameterize the unitary group $SU(2^N)$. The dimension of this
group is $2^{2N}-1$ and there are $2N$ parameters for each vertex.
Thus, we must have $\label{condition} 2N\nu\geq 2^{2N}-1.$
We use $\nu=12$ for the three-qubit gates and $\nu=4$ for the two-qubit
gates. Within this formulation the method of finding the desired
gates is similar to the recently introduced method of
finding holonomic quantum gates \cite{ours}. Thus we again expect
the minimization landscape to be rough and we apply the robust
polytope algorithm \cite{poly} for the minimization.

We concentrate on finding two- and three-qubit gates, since
one-qubit gates can be trivially constructed with the help of
Euler angles. A larger quantum gate could be performed by
factoring it into two- and three-qubit operations, and the
implementation for these could be found numerically. It seems that
quantum operations for four, five or more qubits could be found
with the same method, assuming that sufficient computing resources
are available. However, even in the case of three-qubit gates the
optimization task becomes challenging and we need to use parallel
programming. In the parallel three-qubit program, since the
function evaluations of $f(X_\gamma)$ require a major part of the
computation, we distribute the workload such that each processor
calculates the contribution of a single edge of the polygon. In
addition, one processor handles the minimization routine.

\begin{figure}
\begin{picture}(130,190)
\footnotesize
\put(-50,5){\includegraphics[width=0.42\textwidth]{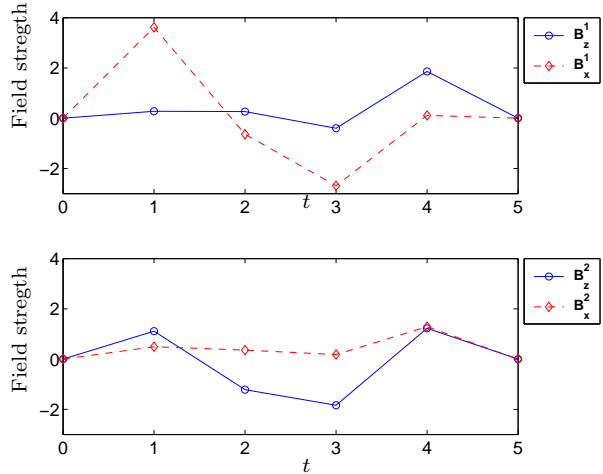}}
\put(50,100){$t$}
\put(50,0){$t$}
\put(-60,30){\begin{sideways} Field stregth\end{sideways}}
\put(-60,130){\begin{sideways} Field stregth\end{sideways}}
\end{picture}
\caption{\label{fg:cnot} Control-parameter sequences as functions
of time that yield the gate in Eq.~(\ref{cnot}) which is
equivalent to the controlled-NOT. The relative error is on the
order of $10^{-11}$ and 100 discretization points per edge were
used.}
\end{figure}

Let us turn to the results. First, we attempt to construct a gate
equivalent to the controlled-NOT, namely
\beq\label{cnot}
U=\exp\left(i\frac{\pi}{4}\right) \left[ \begin{matrix}1&0&0&0
\\0&1&0&0  \\ 0&0&0&1 \\ 0&0&1&0\end{matrix} \right].
\eeq The
phase factor is needed in order for the gate to belong to $SU(4)$.
It is already hard to see from the form of the Hamiltonian how
this gate would be carried out in the present setting.
Figure~\ref{fg:cnot} illustrates an implementation of this gate
that has been obtained by minimizing the error function in
Eq.~(\ref{eq:error}); the error is negligible. This example clearly
illustrates the potential of our method.

\begin{figure}
\begin{picture}(130,190)
\footnotesize
\put(-50,5){\includegraphics[width=0.42\textwidth]{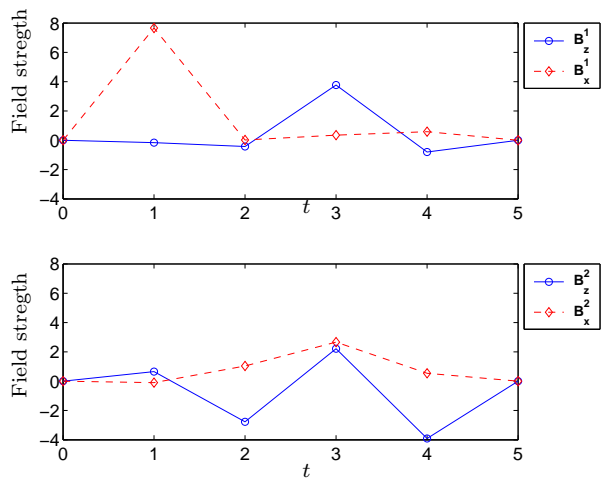}}
\put(50,100){$t$}
\put(50,0){$t$}
\put(-60,30){\begin{sideways} Field stregth\end{sideways}}
\put(-60,130){\begin{sideways} Field stregth\end{sideways}}
\end{picture}
\caption{\label{fg:fourier2} Control-parameter sequences as
functions of time that yield the two-qubit Fourier transform
in Eq.~(\ref{eq:F2}). The relative error is on the order
of $10^{-11}$ and 100 discretization points per edge were used.}
\end{figure}

As a second example, we construct the two-qubit Fourier
transform. The quantum Fourier transform (see e.g.,
Ref.~\cite{gruska}) is given in the case of two qubits by \beq
\label{eq:F2} F_2=\frac{1}{2}\left[\begin{matrix}
1&1&1&1 \\
1&i&-1&-i \\
1&-1&1&-1 \\
1&-i&-1&i
\end{matrix}\right].
\eeq
Furthermore, we need to multiply this by $\exp\left(i\frac{\pi}{8}\right)$
in order to find a gate that belongs to $SU(4)$.
Figure~\ref{fg:fourier2} shows the resulting loop that has been found with the help of the algorithm.
In general, the optimization task for two-qubit gates  can be
performed quite easily with the help of personal computers.
However, finding three-qubit gates is already quite
time-consuming. It proves worth the extra effort to do this, though.

\begin{figure}
\begin{picture}(130,190)
\footnotesize
\put(-50,5){\includegraphics[width=0.45\textwidth]{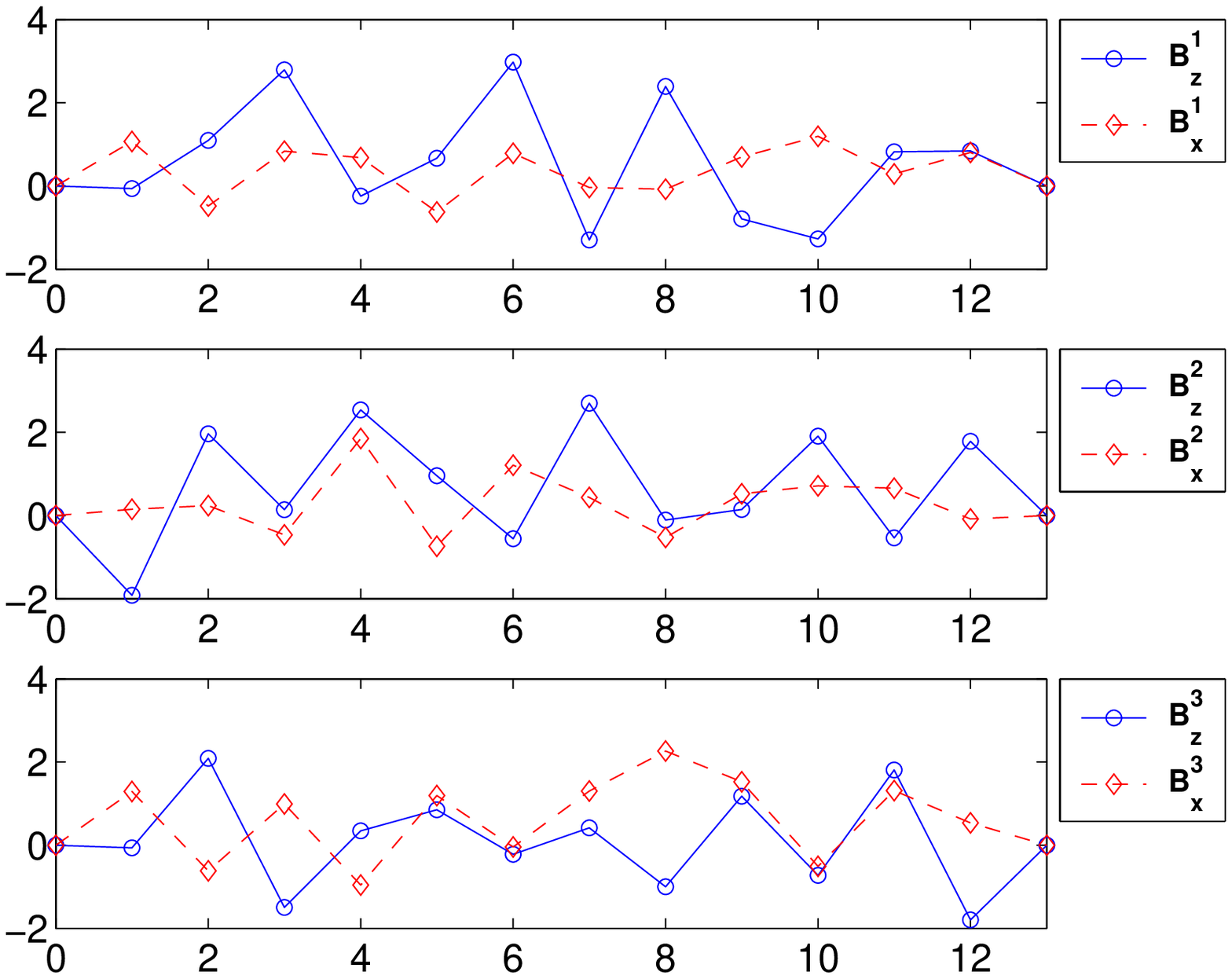}}
\put(60,0){$t$}
\put(60,65){$t$}
\put(60,125){$t$}
\put(-60,70){\begin{sideways} Field stregth\end{sideways}}
\end{picture}
\caption{\label{fg:fourier3} Control-parameter sequences as
functions of time that yield the three-qubit quantum Fourier
transform (modulo a global phase). The relative error is on the
order of $10^{-5}$ and 100 discretization points were used.}
\end{figure}

The three-qubit quantum Fourier transform is \cite{gruska}
\beq
F_3=\frac{1}{\sqrt{8}}\left[ \begin{smallmatrix}
 & & & & & & & \\
1&1&1&1&1&1&1&1\\
1&\omega&\omega^2&\omega^3&\omega^4&\omega^5&\omega^6&\omega^7\\
1&\omega^2&\omega^4&\omega^6&1&\omega^2&\omega^4&\omega^6\\
1&\omega^3&\omega^6&\omega&\omega^4&\omega^7&\omega^2&\omega^5\\
1&\omega^4&1& \omega^4&1&\omega^4&1& \omega^4\\
1&\omega^5&\omega^2&\omega^7&\omega^4&\omega^1&\omega^6&\omega^3\\
1&\omega^6&\omega^4&\omega^2&1&\omega^6&\omega^4&\omega^2\\
1&\omega^7&\omega^6&\omega^5&\omega^4&\omega^3&\omega^2&\omega \\
\end{smallmatrix} \right],
\eeq
where $\omega=\exp\left(i\frac{\pi}{4}\right)$. Since
$\det(F_3)=i$ we must set $
\hat{U}=\exp\left(-i\frac{\pi}{16}\right)F_3 $ such that
$\hat{U}\in SU(8)$. As an evidence of the success of the
three-qubit algorithm, we have in Fig.~\ref{fg:fourier3} plotted
the implementation of the three-qubit Fourier transform.
We conclude from these three examples that it is possible to find
far more powerful optimal implementations of multiqubit quantum
gates with the help of the minimization scheme.

To further assess the strength of the technique, we compare the
number of steps that are required to carry out the three-qubit
Fourier transform using only two-qubit gates with the number of
steps required when using the full three-qubit implementation of
Fig.~\ref{fg:fourier3}. The two-qubit implementation~\footnote{We
assume that one-qubit operations are embedded into two-qubit
gates.} requires effectively four gates, see
Fig.~\ref{fg:principle}. Since these gates would have to be
performed sequentially, we would need five polygon edges per
two-qubit operation. This results in 20 edges for the whole
operation. Using elementary gates would require far more edges.
Our optimized three-qubit Fourier transform, though, only requires
13 edges. Since each edge contributes the same amount to the
operation time, we conclude that our implementation is improved.
What is more, not all multiqubit gates can be decomposed as
conveniently as the Fourier transform. For them the gain is
higher. Thus, increasing the amount of classical computing
resources should yield even better results.

In conclusion, we have described how to efficiently construct two-
and three-qubit quantum gates for the Josephson charge qubit using
numerical optimization. An immediate strength of the present
scenario is that one avoids unnecessary idle time during the
logical quantum operations. Since the loops are traversed at a
piecewise constant speed, and no fields are instantaneously
switched, this method of constructing quantum gates should be
viable from the experimental point of view as well. The effect of
finite fall and rise times of pulses on the quality of quantum
gates has been studied recently \cite{oh}. Since we do not use
pulses but instead interpolate along linear paths in the parameter
space, such errors can be avoided. It seems reasonable to
construct large-scale quantum algorithms in multiqubit blocks.
This can be accomplished by optimizing the gate realization with the
help of classical computers.

The authors would like to thank M. Nakahara for useful discussions.
AON and JJV thank the Research Foundation of Helsinki University of Technology
and the Graduate School in Technical Physics for financial support;
this work is supported by the Academy of Finland through a Research Grant in Theoretical Materials Physics.
We also thank the Center for Scientific Computing (CSC, Finland) for computing resources.


\begin{thebibliography}{99}

\bibitem{elementary}
A. Barenco, C. H. Bennett, R. Cleve, D. P. DiVincenzo,
N. Margolus, P. Shor, T. Sleator, J. Smolin, and H. Weinfurter,
Phys. Rev. A {\bf 52}, 3457 (1995).

\bibitem{gruska}
J. Gruska, \emph{Quantum Computing}, McGraw-Hill, New York (1999);
 M. A. Nielsen and I.L. Chuang, \emph{Quantum Computation and Quantum Information},
Cambridge University Press, Cambridge (2000); A. Galindo and M. A. Martin-Delgado,
Rev. Mod. Phys. {\bf 74}, 347 (2002).

\bibitem{averin}
A. Shnirman, G. Schön, and Z. Hermon, Phys. Rev. Lett. {\bf 79}, 2371 (1997);
D. V. Averin, Solid State Commun. {\bf 105}, 659 (1998).

\bibitem{schon}
Yu. Makhlin, G. Sch\"{o}n, and A. Shnirman,
Rev. Mod. Phys. {\bf 73}, 357 (2001).

\bibitem{You}
J. Q. You, J. S. Tsai, and F. Nori, Phys. Rev. Lett. {\bf 89}, 197902 (2002).

\bibitem{zurek}
W. H. Zurek, Rev. Mod. Phys. {\bf 75} (in print, 2003).

\bibitem{nakamura}
Y. Nakamura, Yu. A. Pashkin, and J. S. Tsai,
Nature {\bf 398}, 786 (1999);
Phys. Rev. Lett. {\bf 87}, 246601 (2001).

\bibitem{pc}
T. P. Orlando, J. E. Mooij, L. Tian, C. H. van der Wal, L.
Levitov, S. Lloyd, and J. J. Mazo, Phys. Rev. B {\bf 60}, 15398
(1999).

\bibitem{martinis}
J. M. Martinis, S. Nam, J. Aumentado, and C. Urbina,
Phys. Rev. Lett. {\bf 89}, 117901 (2002); Y. Yu, S. Han, X. Chu, S. Chu,
and Z. Wang, Science {\bf 296}, 889 (2002).

\bibitem{vion}
D. Vion, A. Aassime, A. Cottet, P. Joyez, H. Pothier, C. Urbina,
D. Esteve, and M. H. Devoret, Science {\bf 296}, 886 (2002).

\bibitem{ours}
A. O. Niskanen, M. Nakahara, and M. M. Salomaa, Quantum Information and
Computation {\bf 2}, 560 (2002); Phys. Rev. A {\bf 67}, 012319
(2003).

\bibitem{zanardi}
P. Zanardi and M. Rasetti,
Phys. Lett. A {\bf 264}, 94 (1999).

\bibitem{siewert}
J. Siewert and R. Fazio,
Phys. Rev. Lett. {\bf 87}, 257905 (2001).

\bibitem{palao}
J. P. Palao and R. Kosloff,
Phys. Rev. Lett. {\bf 89}, 188301 (2002).

\bibitem{wang}
X. Wang, A. S\o rensen, and K. M\o lmer, Phys. Rev. Lett. {\bf
86}, 3907 (2001).

\bibitem{poly}
J. C. Lagarias, J. A. Reeds, M. H. Wright, and P. E. Wright,
SIAM J. Optim. {\bf 9}, 112 (1998).

\bibitem{oh}
S. Oh,
Phys. Rev. B {\bf 65}, 144526 (2002).

\end{thebibliography}
\end{document}